\documentclass{aa}
\usepackage{txfonts}
\usepackage{graphicx}
\usepackage{natbib}
\bibpunct{(}{)}{;}{a}{}{,}
\begin{document}
  
\title{Photodesorption of water ice}
\subtitle{A molecular dynamics study}
  
\author{S. Andersson\inst{1,2,3} \and E. F. van Dishoeck\inst{1}}

\institute{Leiden Observatory, Leiden University, P.O. Box 9513, 2300
RA Leiden, The Netherlands \and Gorlaeus Laboratories, Leiden Institute of Chemistry,
Leiden University, P.O. Box 9502, 2300 RA Leiden, The Netherlands \and 
Department of Chemistry, Physical Chemistry, Universtity of Gothenburg, 
SE-41296 Gothenburg, Sweden}

\date{Received date /Accepted date}

\abstract {Absorption of ultraviolet radiation by water ice 
coating interstellar grains can lead to dissociation and desorption of
the ice molecules. These processes are thought to be important in the
gas-grain chemistry in molecular clouds and protoplanetary disks, but
very few quantitative studies exist.}  
{We compute the photodesorption
efficiencies of amorphous water ice and elucidate the mechanisms by
which desorption occurs.}  
{Classical molecular dynamics calculations
were performed for a compact amorphous ice surface at 10 K
thought to be representative of interstellar ice. Dissociation and
desorption of H$_2$O molecules in the top six monolayers are considered
following absorption into the first excited electronic state with
photons in the 1300--1500 \AA\ range. The trajectories of the H and OH
photofragments are followed until they escape or become trapped in the
ice.}  
{The probability for H$_2$O desorption per absorbed UV photon
is 0.5--1\% in the top three monolayers, then decreases to 0.03\% in the
next two monolayers, and is negligible deeper into the ice.
The main H$_2$O removal mechanism
in the top two monolayers is through separate desorption of H and OH fragments.
Removal of H$_2$O molecules from the ice, either as H$_2$O itself or its products,
has a total probability of 2--3\%
per absorbed UV photon in the top two monolayers. In the third monolayer the probability is
about 1\% and deeper into the ice the probability of photodesorption falling to insignificant numbers.  
The probability of any removal of H$_2$O per incident photon is estimated to be $3.7 \times 10^{-4}$,
with the probability for photodesorption of intact H$_2$O molecules being $1.4 \times 10^{-4}$
per incident photon.
When no desorption occurs, the H and OH products can travel up to 70 and 60
\AA\ inside or on top of the surface, respectively, during which they can react with
other species, such as CO, before they become trapped.
} 
{}

\keywords{astrochemistry -- molecular data -- ISM: molecules}

\maketitle

\section{Introduction}

Ices are a major reservoir of the heavy elements in a variety of
astrophysical environments, ranging from cold and dense molecular
clouds \citep[e.g.,][]{wil82,whi88,smi89,mur00,pon04} and protoplanetary disks
\citep{pon05,ter07}
to the icy bodies in our own solar system such as comets
\citep[e.g.,][]{mum93} and Kuiper Belt Objects 
\citep{jew04}. In star-forming
clouds, the fraction of carbon and oxygen locked up in ice is
comparable to that in gaseous CO \citep{pon06,vand96}, 
whereas at the centers of cold pre-stellar cores more than 90\% of the
heavy elements can be frozen out \citep[e.g.,][]{cas99,ber02}.
Similarly, the cold midplanes of protoplanetary disks
around young stars are largely devoid of gaseous molecules other than
H$_2$, H$_3^+$ and their isotopologues \citep[e.g.,][]{aik02,cec05}. 
Thus, a good understanding of how molecules adsorb
and desorb from the grains is critical to describe the chemistry in
regions in which stars and planets are forming.

The importance of ultraviolet (UV) radiation in affecting interstellar
ices is heavily debated in the literature. On the one hand, the large
extinctions of 100 mag or more along the lines of sight where ices are
detected prevent UV radiation from penetrating deep into the clouds \citep[e.g.,][]{ehr01,sta04}, unless there are
cavities through which the stellar UV photons can escape \citep{spa95}. Thus,
the bulk of the ices are thought to be shielded from both external and
internal radiation sources in which case photodesorption is thought to be
unimportant \citep[e.g.,][]{leg85,har90}. On the other
hand, UV photons are produced locally throughout the cloud by the
interaction of cosmic rays with the gas, albeit at a level about
$10^4$ times less than that of the general interstellar radiation
field \citep{pra83,she04}. In addition,
X-rays from young stars penetrate much further into their surroundings
than UV and can produce local UV photons through a similar process
\citep{dal99,sta05}.  Moreover, the
observed emission of optically thick millimeter lines from gaseous
molecules is often dominated by the outer layers of the cloud where UV
photons play a role.  These UV photons can be important not only in
the desorption of ices but also in the creation of reactive
photo-products such as energetic H atoms and radicals which can move
through the ice and encounter other species leading to the formation
of more complex molecules \citep[e.g.,][]{dhe82,gar06}.

Water ice is the dominant consituent of interstellar ices \citep[e.g.,][]{gib00,pon06}
with an abundance at least three orders of magnitude larger than that
of gaseous water in cold clouds \citep{sne00,boo03}.  Thus,
evaporation of water ice, even at a low fraction, can significantly
affect the gaseous water abundance. Recent models of translucent and
dense clouds invoke photodesorption of water ice in the outer regions
to explain the gaseous water emission observed by the Submillimeter
Wave Astronomy Satellite (SWAS) \citep{ber05}.
Photodesorption is also used to interpret the tentative
detections of HDO and other gaseous species in the surface layers of
protoplanetary disks \citep{wil00,dom05}.
The adopted desorption efficiencies in these models, about 0.1\% per
incident UV photon, are based on a single experiment by
\citet{wes95a,wes95b} exposing ices at 35--100 K to Lyman-$\alpha$ radiation.
The authors did not detect any water photodesorption in the limit of low UV
photon fluence (integrated flux). Therefore it was suggested that water photodesorption
only occurs from ices
that have been subject to large doses of UV photons and not directly upon the first
exposure to UV radiation. This is in contrast with recent experiments reported by \"Oberg
et al. (submitted to ApJ), where there is a clear component of the water photodesorbed from 
amorphous ices at 18--100 K that
is detected directly upon the first exposure to a UV lamp. Apart from that the results
remain quite similar to the ones by Westley et al.

Other experiments on UV irradiation of water ices have also been performed.
\citet{gho71} observed H, OH, and H$_2$O$_2$ following UV irradiation
of crystalline ice at 263 K, while \citet{ger96} found production of OH, HO$_2$, and
H$_2$O$_2$ in the ice upon exposing amorphous ice at 10 K to UV light covering
mainly the first and second electronic absorption bands of H$_2$O. \citet{wat00}
irradiated amorphous D$_2$O ice at 12 K with UV photons
and observed substantial amounts of D$_2$ after
irradiation at $\lambda$ = 126 nm, but very little at $\lambda$ = 172 nm.
In the experiments by \citet{yab06} H atoms were found to desorb from the ice
after UV irradiation at $\lambda$ = 157 nm and $\lambda$ = 193 nm.
There are also a few reports on two- and multi-photon excitation of water ice leading
to photodesorption of H$_2$O molecules \citep{nis84,ber06}. In these cases
the photon energies are below the threshold for absorption in the ice, but
upon multiple absorptions the excitation energies fall between 9 and 10 eV, between
the first and second absorption band in ice.
Clearly, there is a need for more quantitative information
on the processes induced by UV photons in ices, even for the simplest
cases such as pure water ice.

We present here the results of the first theoretical study of the
dissociation of H$_2$O molecules in pure water ice following
absorption by UV photons. In addition to providing probabilities for
desorption to be used in astrochemical models, these simulations
provide insight into the mechanisms leading to desorption as well as
the movements of the energetic photoproducts in the ice before they
become trapped. In Sect.\ 2, we present the methods used in this
study, in Sect.\ 3 the main results, and in Sect.\ 4 a short
discussion and astrochemical implications. In Sect.\ 5 the results are summarized
and some concluding remarks are given.

\section{Method}

All our calculations have been performed using classical Molecular
Dynamics (MD) methods \citep{all87} with analytical
potentials. Details of the computational procedure have been described
in \citet{and06}; here only a brief outline of
the methods will be presented. 

\subsection{Amorphous water ice}

To create an amorphous ice slab, the procedure outlined in \citet{alh04a}
was used. In brief, a slab of 8 bilayers (16
monolayers) of crystalline ice was first created consisting of a cell
containing 480 H$_2$O molecules. The cell has the dimensions $x$: 22.4
{\AA}, $y$: 23.5 {\AA}, and $z$: 29.3 {\AA}.  Periodic boundary
conditions are applied in the $x$- and $y$-directions, the $z$
coordinate being parallel to the surface normal. Thus an infinite ice
surface is created. The H$_2$O molecules are treated as rigid rotors
and their interactions are governed by the TIP4P potential \citep{jor83},
which describes the interaction as a sum
of pair interactions (electrostatic and Lennard-Jones potentials). The
two bottom bilayers are kept fixed to simulate bulk ice and the
molecules in the other 6 bilayers are allowed to move without any
dynamical constraints other than that they remain rigid. The dynamics
are at all times governed by classical Newtonian mechanics. To force
the transition to amorphous ice, the surface is initially allowed to
equilibrate for 5 ps at 10 K but then the temperature is increased to
300 K using a computational equivalent of a thermostat \citep{ber84}. 
In this way the top bilayers form a
liquid. The system is left to equilibrate for 100 ps after which it is
rapidly cooled to 10 K. Then it is once again equilibrated for 100
ps. The resulting amorphous ice structure most closely resembles the
structure of compact amorphous ice obtained
experimentally and is thought to be representative of the
structure of interstellar water ice \citep{alh04a,alh04b}.
See Fig.\ 4 of \citet{alh04a} and Fig.\ 2 of \citet{and06} for images.
It does
not exhibit the microporous structure that is obtained in vapor
deposited ice \citep{may86,kim01a,kim01b}.
Given the dimensions of the simulation cell
such a structure is simply not possible to obtain, since the pores should have a size on
the same order as the cell we use.
However, we believe that it
is a good representation of an amorphous ice surface on a local scale, whether that
be at the ``outside" of the surface or inside a void deeper in the ice.
Implications of this for the obtained results are discussed in Sect. \ref{DiscAstro}.

In the rest of the paper we will discuss the depth into the ice in terms
of {\em monolayers}. To avoid confusion our definition of monolayer is
the thickness of ice corresponding to half a crystalline bilayer, i.e., if
the ice were crystalline each bilayer would consist of two monolayers.
For ease of definition the monolayers have been taken to be divided according
to the $z$ values of the centers-of-mass of the molecules, e.g., the 30 molecules
with the largest values of $z$ constitute the top (``first") monolayer.

\subsection{Initial conditions}

Once the ice surface is set up, one H$_2$O molecule is chosen to be
photodissociated. This molecule is then made completely flexible and
its intramolecular (internal) interactions are governed by an analytic
potential energy surface (PES) for the first electronically excited state
(the \~A$^1{\mathrm B}_1$ state) of
gas-phase H$_2$O based on high-quality \emph{ab initio} electronic
structure calculations \citep{dob97}. This excited
potential is fully repulsive so that absorption into this state leads
to dissociation of the H$_2$O molecule into H + OH. The intermolecular
interactions of the excited state H$_2$O with the surrounding H$_2$O
molecules are governed by specially devised partial charges for
describing the electrostatic interactions.
In short, a charge of -0.2$e$ is put on the O atom and charges of
+0.1$e$ on the H atoms. This gives a smaller dipole moment than
that of the ground state H$_2$O potential. The effect of this is
that a less favorable interaction is obtained with the surrounding H$_2$O
molecules, giving higher excitation energies than with the uncorrected
gas-phase potential energy surface. This leads to the blue-shift of about
1 eV of
the ice UV spectra seen in Fig. \ref{FigSpect}, which
agree well with the first UV absorption band in amorphous and
crystalline ices. If the ground state partial charges
were to be used for the excited state the excitation spectrum would
coincide with the gas-phase UV spectrum.
For more
details on the potentials see \citet{and06}.
Similar procedures using potential energy surfaces for higher
excited states (\~B, \~C, etc.) should in principle give reasonable
representations of higher-lying absorption bands in water ice.

The initial internal
coordinates and momenta of the atoms in the selected molecule are
sampled by a Monte Carlo procedure using a semi-classical (Wigner)
phase-space distribution \citep{sch93} that has been
fitted to the ground-state vibrational wave function of H$_2$O 
\citep{vanh01}. This procedure gives initial
conditions that are very similar to those found in fully quantum
mechanical methods and has been shown to work well for the description
of photodissociation processes of gaseous molecules. The transition
dipole moment function, which governs the strength of the absorption,
is taken from the calculation for gaseous H$_2$O by \citet{vanh00}.

Dissociation of molecules in the top six monolayers has been
considered. For each monolayer all 30 molecules have been
dissociated, one molecule at a time.
For each molecule 200 configurations and momenta were
sampled from the Wigner distribution. This gives 6000 trajectories per
monolayer and 36000 trajectories in total.

\subsection{Calculation of spectra}

The excitation energy is computed by taking the energy difference
between an ice slab with an excited state H$_2$O and one with a ground
state H$_2$O (with the same coordinates).  Each excitation is assigned
a weight calculated as the square of the coordinate-dependent
transition moment.  By summing the weights of the excitation energies
binned in 0.05 eV-wide energy intervals, ``intensities" are
obtained. Taken together these intensities form a UV absorption
spectrum for the ice.  The monolayers 5 to 6 were found to be
converged to a ``bulk behavior" \citep{and05}
and could therefore be used to compare the calculated spectra to
experimental data. The gas-phase spectrum presented in Sect.\
\ref{ResDisc} was obtained using the same intramolecular potential
surfaces as above, but without the surrounding molecules and with 1000
sampled configurations.

\subsection{Dynamics of the dissociating molecule}

After putting the molecule in the excited state, the dissociating
trajectory is integrated with a timestep of 0.02 fs. A maximum time of
20 ps has been used before terminating the trajectory. Most of the
trajectories (99.6\%) were terminated before that because the system was found
in one of the final outcomes (see Sect. \ref{ResDisc}) with
negligible probability of transforming into a different state.

When the excited H$_2$O dissociates, the intermolecular interactions
are smoothly switched into separate interactions between the
photoproducts and water ice, i.e., H-H$_2$O and OH-H$_2$O
potentials. All details of the potentials and the functions used to switch between
different potentials are given in \citet{and06}. The switching functions connect
the partial charges, the dispersion interactions and repulsive potentials between
the H$_2$O-H$_2$O potentials and the OH-H$_2$O and H-H$_2$O potentials affecting the
dissociating molecule. The switches are functions of the OH distances ($R_{\mathrm{OH}}$) within this molecule
and will give the interaction parameters as continuous functions in the range 1.1 --
1.6 \AA\ in $R_{\mathrm{OH}}$. The intramolecular potential is switched to the
ground-state PES, which allows H and OH to recombine to form H$_2$O. This
switch is done in an analogous way as for the intermolecular interactions above, but here the
range of $R_{\mathrm{OH}}$ where the switch is made is 3.0 -- 3.5 \AA. In this range
the excited-state and ground-state PES are near-degenerate, so a high transition probability
between the two states is quite probable. Once $R_{\mathrm{OH}}$ becomes larger than 3.5 \AA\ the system will remain
on the ground-state PES, even if $R_{\mathrm{OH}}$ again becomes smaller than 3.5 \AA. This is what allows for recombination
of H and OH.
The intermolecular interactions for the recombined ground state H$_2$O with the
surrounding H$_2$O molecules are taken from the TIP3P potential
\citep{jor83}.

A slightly different stop criterion has been used in these calculations compared to the results presented
previously \citep{and05,and06}. When an H atom or OH is accommodated to the
ice surface (``trapped") the trajectory is run until its translational energy equals $k_{\rm B} T$ or lower
{\em and} the binding energy to the surface is 0.02 eV (H atom) or 0.1 eV (OH) or stronger. In the older
version of the code, the stop criterion was based on the kinetic energy of the individual {\em atoms}
in relation to the potential energy \citep{and06}. The introduction of the new termination
scheme led to a reduction by about 50\% of the number of trajectories exceeding 20 ps.

Although most of the results presented here focus on amorphous ice,
calculations have been performed for crystalline ice as well for
comparison. Details can be found in \citet{and06}.

\section{Results and Discussion}
\label{ResDisc}

\subsection{Ice UV absorption spectrum}

\begin{figure}[h]
\resizebox{\hsize}{!}{\rotatebox{270}{{\includegraphics{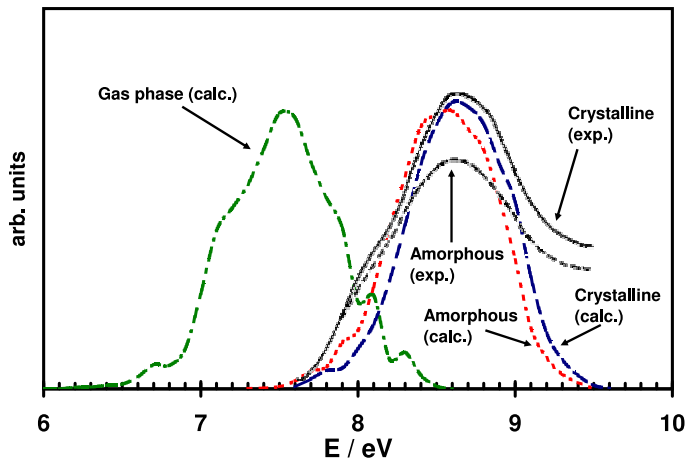}}}}
   \caption{Calculated and experimental \citep{kob83} spectra of
   the first UV absorption band in crystalline and amorphous ices and
   calculated first absorption band for gas-phase H$_2$O. Note that
   the intensity is given in arbitrary units and that the intensities
   of the ice spectra have been scaled to roughly coincide.}
   \label{FigSpect}
\end{figure}

As presented in Fig. \ref{FigSpect}, our calculated spectra of the
first UV absorption bands in amorphous and crystalline ice match very
well the experimentally obtained spectra, both in general shape as
well as in the peak and threshold energies. The calculated gas-phase
spectrum of the first absorption band shown in Fig. \ref{FigSpect}
also matches the experimental peak energy (7.4--7.5 eV) quite well.
This is naturally to be expected, since the potential surfaces used
are based on very high-quality \emph{ab initio} calculations of the
energy. The success in reproducing the measured spectra leads us to
believe that the amount of excess energy released into the ice is
basically correct.

The ice spectrum is blue-shifted with respect to that of gaseous
H$_2$O and has significant cross section only in the 7.5--9.5 eV
range. Thus, the photodesorption probabilities computed here are
appropriate for photons in the 1300--1500 \AA\ range.
Dissociation of H$_2$O can also occur following absorption into
higher excited states (e.g., the equivalent of the \~{B} state
of gaseous H$_2$O) but these generally contribute less
than 20\% of the total absorption in a dense cloud.

\subsection{Photoprocesses and desorption probabilities}
\label{PhotoDesProb}

\subsubsection{Overall probabilities}
\label{OverProb}

\begin{figure}[h]
\resizebox{\hsize}{!}{\rotatebox{270}{{\includegraphics{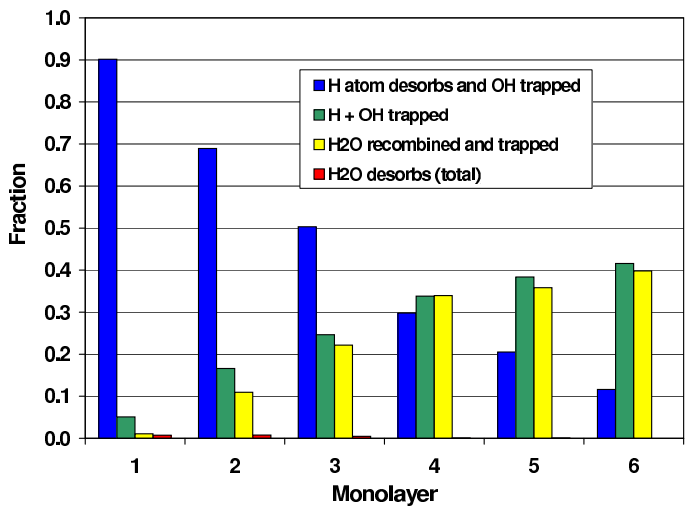}}}}
   \caption{Fractions of the main outcomes after H$_2$O photodissociation for the top six monolayers of amorphous ice. These probabilities
            are calculated from all trajectories, irrespective of excitation energy.}
    \label{FigMainOut}
\end{figure}

Photodissociation of a water ice molecule can have several outcomes,
with the H and OH photoproducts either becoming trapped in the ice,
recombining back to an H$_2$O molecule, or desorbing from the ice
surface.  Fig. \ref{FigMainOut} shows the probabilities (as fractions
of the number of absorbed photons) of the main processes as functions
of how deep into the ice the dissociating molecule initially is located. 
Note that these probabilities are
given per {\em absorbed} UV photons and {\em not} per incident photon.
In the first monolayer the dominant outcome is
that the hydrogen atom desorbs and the OH radical is
trapped in or on the ice. The probability of this event drops rapidly
from about 0.9 in the first monolayer to just over 0.1 in the sixth
monolayer. This reflects the effect of the ice on the motion of the H
atom. At the surface there are very few molecules to stop the H atom
from desorbing, but starting from deeper into the ice there are more
obstacles on the way to the gas phase.

For the same reasons the probabilities of the other two major outcomes
steadily increase as one moves deeper into
the ice. These are the events when either both
H and OH become separately trapped in the ice or when H and OH
recombine to form H$_2$O to subsequently become trapped in the
ice. Except for the top two monolayers the probabilities of these two
events are roughly equal and in the sixth monolayer the probabilities
are up to 0.4. The reason for the lower probability of the H$_2$O
molecule being recombined and trapped in the top monolayers can be
understood from the open structure of the uppermost layers, which more
easily allows for the photofragments to escape the region of the ice
where they were initially formed. 

\begin{table}
\caption{Total probabilities of H atom, OH, and H$_2$O desorption (per absorbed UV photon) as functions of monolayer}
\label{TabDes1}
\centering
\begin{tabular}{c c c c}
\hline\hline
ML & H atoms & OH & H$_2$O \\
\hline
 1 & 0.92 & 0.024 & 7.3$\times10^{-3}$ \\
 2 & 0.70 & 0.015 & 7.7$\times10^{-3}$ \\
 3 & 0.51 & 4.0$\times10^{-3}$ & 4.8$\times10^{-3}$ \\
 4 & 0.30 & 0.00 & 3$\times10^{-4}$ \\
 5 & 0.21 & 0.00 & 3$\times10^{-4}$  \\
 6 & 0.12 & 0.00 & 0.00 \\
\hline
\end{tabular}
\end{table}

The probability of photodesorption of H$_2$O is seen to be low
compared with the above processes, 0.7\% in the top layer and 0.8\% in the second layer, and
then decreases with distance from the surface (see Table \ref{TabDes1}).
However, this is only considering the desorption of {\em intact} H$_2$O molecules.
If one is interested in the removal of H$_2$O from the surface without considering what enters
the gas phase, the dominant mechanism in the top two monolayers is actually desorption
of separate H and OH fragments. Desorption of OH is in most cases accompanied by the
desorption of an H atom, but a minor fraction of OH desorption occurs with the H atom being 
trapped in the ice (see Sect. \ref{MechProb}).
In summary, the desorption probabilities of OH and H$_2$O are about 2 orders of magnitude
lower than that of H atoms with OH desorption being about twice as probable as H$_2$O desorption
if one sums over the probabilities from all monolayers (see also Sect. \ref{DiscAstro}).

\subsubsection{Mechanisms and their probabilities}
\label{MechProb}

\begin{figure}[h]
\resizebox{\hsize}{!}{\includegraphics{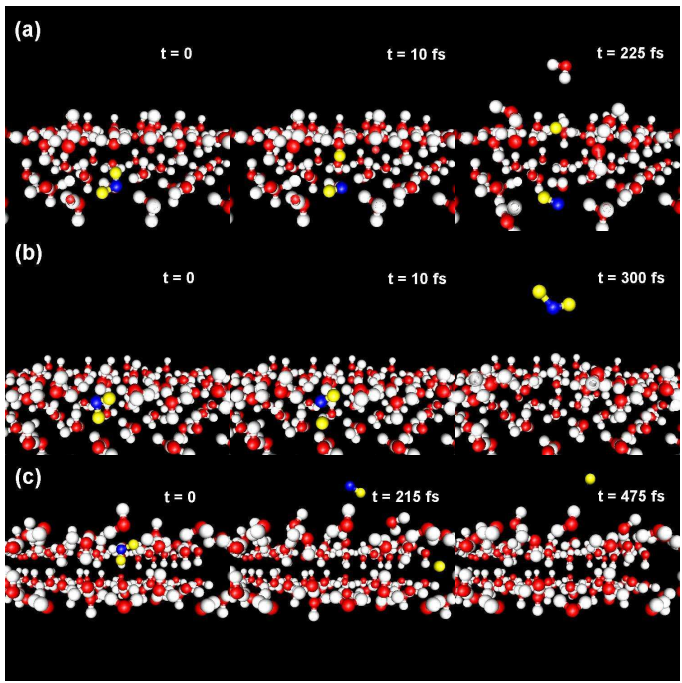}}
   \caption{Snapshots of trajectories of mechanisms of H$_2$O
   desorption for a crystalline ice model. (a) One of the surrounding
   molecules desorbs. (b) The photofragments H and OH recombine and
   desorb as H$_2$O. (c) The photofragments both desorb as separate
   species. The red and white atoms correspond to O and H atoms in the
   surrounding molecules and the blue and yellow atoms correspond to
   the O and H atoms of the photodissociated H$_2$O molecule.}
   \label{FigMovie}
\end{figure}

Analysis of the
trajectories shows that there are three distinct mechanisms
for H$_2$O removal (see Fig. \ref{FigMovie}). Note that these
snapshots are taken from calculations on crystalline ice
\citep{and06} for ease of visualization: (a) An H atom
released from photodissociation of H$_2$O is able to transfer enough
momentum to one of the other H$_2$O molecules to ``kick" it off the
surface, (b) H and OH recombine to form H$_2$O and subsequently desorb, and (c) the H
and OH both desorb from the surface separately.
In Table
\ref{TabDes2} the absolute probability of H$_2$O desorption is given
along with the relative probabilities of the three different
mechanisms for each monolayer. With the additional mechanism (c) the
probability of H$_2$O removal is 2.7\% and 1.9\% per absorbed UV photon in the
first and second monolayer, respectively.

\begin{table}
\caption{Absolute probabilities for removal of H$_2$O from an amorphous ice 
surface (per absorbed UV photon) as functions of monolayer
}
\label{TabDes2}
\centering
\begin{tabular}{c c c c c}
\hline\hline
ML & H$_2$O$^a$ & (a) & (b) & (c) \\
 & desorption & H$_2$O& H$_2$O& H + OH \\
 & probability & intact & intact & fragments \\
\hline
 1 & 0.027 & 0.10 & 0.17 & 0.73 \\
 2 & 0.019 & 0.16 & 0.24 & 0.60 \\
 3 & 7.0$\times10^{-3}$ & 0.43 & 0.26 & 0.31 \\
 4 & 3$\times10^{-4}$ & 1.00 & 0.00 & 0.00 \\
 5 & 3$\times10^{-4}$ & 1.00 & 0.00 & 0.00 \\
\hline
\multicolumn{5}{l}{$^a$Probability for desorption as H$_2$O or H + OH.}
\end{tabular}
\end{table}

\begin{figure}[h]
   \caption{Fractions of the detailed outcomes after H$_2$O
   photodissociation for the top five monolayers of amorphous ice. The
   error bars correspond to a 95\% confidence interval.}
\resizebox{\hsize}{!}{\rotatebox{270}{{\includegraphics{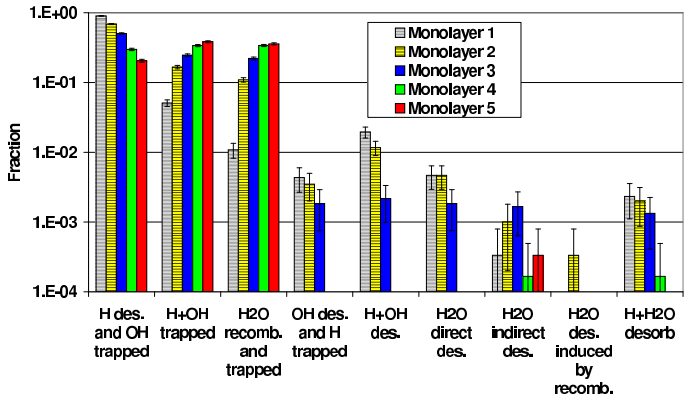}}}}
   \label{FigDetOut}
\end{figure}

In the top two monolayers the direct desorption of H and OH fragments is the dominant desorption
mechanism, but in the third layer the three distinct desorption mechanisms (``a", ``b", and ``c") are
roughly equally probable.
Following UV absorption in monolayers 4 and 5 only the indirect ``kick-out"
mechanism is effective. In Fig. \ref{FigDetOut} the probabilities of all mechanisms of
desorption of H atoms, OH radicals, and H$_2$O molecules are presented for the top five monolayers.
The desorption of OH is possible either together with the H atom as shown above or separately with
the H atom remaining trapped. The former mechanism is found to have a higher probability. In total,
the desorption of OH is about a factor of 2 more probable than the desorption of H$_2$O. Below the
third monolayer the released OH radicals do not have sufficient kinetic energy to make it to the top of the
surface {\em and} desorb. The rightmost
three categories constitute a further division of the category of indirect H$_2$O desorption.
Here ``H$_2$O indirect desorption" only refers to the cases where a molecule is kicked out by an H atom, which
subsequently remains in the ice. The category ``H$_2$O desorption induced by recombination" refers to the rare occurence
where it is the excess energy from the recombination of the H and OH fragments that kicks the molecule off the surface.
The case ``H + H$_2$O desorb" refers to when the H atom kicks the molecule off the surface and subsequently desorbs
itself. This last category dominates the indirect desorption in the first two monolayers and
also consitutes the maximum amount of matter that has been observed to desorb following
photoexcitation. 
For H$_2$O photoexcited below the fifth
monolayer there is no evidence of photodesorption of H$_2$O molecules.

The division into monolayers in Table \ref{TabDes2} refers to where the
photoexcited H$_2$O is situated.
The H$_2$O molecules that actually desorb upon being expelled by an H atom or recombining H$_2$O molecule
all originate in monolayers 1 (84\%) and 2 (16\%). In most of these cases it is not only the transfer
of momentum that is effective, but also the repulsive interaction from the photoexcited molecule,
which most often is in the near vicinity of the desorbing molecule. To illustrate this one can consider the lowering
in binding energy of the molecule about to be desorbed. In monolayer 1 the average binding energy of all molecules is
0.9 eV (calculated using the TIP4P potential) and of the desorbing molecules prior to excitation it is 0.8 eV. 
However, the photoexcitation lowers the binding
energy of these molecules by on average 0.3 eV. About 25\% of the desorbing molecules do not have their binding
energy significantly lowered by excitation, but are kicked out solely by momentum transfer. 
If these molecules are excluded then the binding energy is on average lowered by 0.4 eV. In the second monolayer only about 10\%
of the desorbing molecules do not have their binding energy significantly lowered. There the average binding energy
is 1.1 eV and this is lowered by 0.3 eV on average upon photoexcitation.

There is no sign of molecules being electronically excited and then desorbing intact {\em directly}, i.e.,
expelled by the repulsive interaction of the excited state molecule with its surroundings.   
The molecules dissociate very quickly (on the order of 10 fs) and that is not sufficient time for the molecule to desorb
before it is dissociated. As discussed above the photofragments can however recombine and {\em then}
subsequently desorb as H$_2$O.

\begin{figure}[h]
\resizebox{\hsize}{!}{\includegraphics{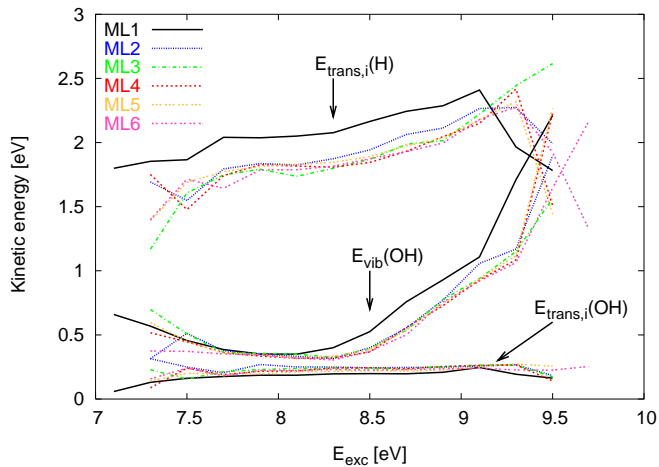}}
    \caption{The average initial translational energies of H atoms [$E_{\mathrm{trans,i}}$(H)] and OH radicals [$E_{\mathrm{trans,i}}$(OH)] and
     the average OH vibrational energy following photodissociation as function of excitation energy
     for the top six monolayers in amorphous ice.}
    \label{FigEtvvEexc}
\end{figure}

\subsubsection{Effects of product energies}
\label{EffProdEn}

The H atoms that are released have an average energy of 1.5-2.5 eV depending on the excitation energy
and to a lesser extent in which monolayer they originate (see Fig. \ref{FigEtvvEexc}). As can be seen the
average initial H atom translational energy increases with increasing excitation energy, but above an excitation
energy of 9 eV it drops to
somewhat lower energies. The average vibrational energy of OH has a minimum value of 0.3 eV around $E_{\mathrm{exc}}$ = 8 eV
but increases strongly to about 2 eV at $E_{\mathrm{exc}}$ = 9.5 eV. This implies that at lower excitation energies 
the vast majority of the 
OH molecules are formed in the vibrational ground state, since the experimental zero-point energy of
OH is 0.23 eV \citep{hub79} (see also \citet{and06}). When the excitation energy is increased above 9 eV,
large fractions of vibrationally excited OH are produced. 
The average initial translational energy of OH is only weakly
dependent on excitation energy and lies around 0.2 eV with only a slight increase with increasing $E_{\mathrm{exc}}$.

\begin{figure}[h]
\resizebox{\hsize}{!}{\includegraphics{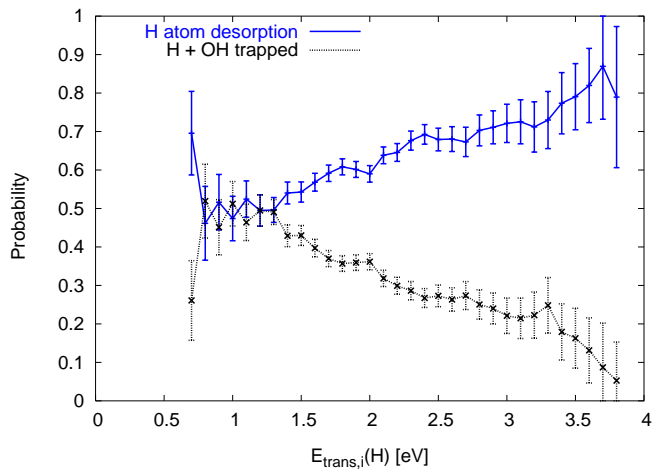}}
    \caption{The probabilities of H atom desorption and trapping of H and OH as function of initial
    H atom translational energy averaged over the top six monolayers in amorphous ice.
    Note that for technical reasons these are probabilities with the possibility of H$_2$O recombination
    excluded (see text). The error bars correspond to a 95\% confidence interval.}
    \label{FigOutvEtH}
\end{figure}

In Fig. \ref{FigOutvEtH} the H atom desorption probability is plotted alongside the probability of H and OH both becoming
trapped as functions of initial H atom translational energy.
The plotted probabilities are somewhat higher than they should be because
the probability of recombination of H$_2$O has been excluded in the set of outcomes. The reason for this
is that during recombination the translational energy of the H atom becomes very high. If recombination occurs immediately
after dissociation it is quite difficult to distinguish the maximum translational energy the H atom normally would have
after photodissociation and the maximum translational energy it gets during recombination. However, the trend
is clear that the H atom desorption probability increases with increasing initial translational energy, as one would
intuitively expect. The reason for the unexpectedly high desorption probability at $E_{\mathrm{trans,i}}$(H) = 0.7 eV is not quite 
clear and it could simply be an effect of insufficient sample size, given that the error bars are fairly large.

\begin{figure}[h]
\resizebox{\hsize}{!}{\includegraphics{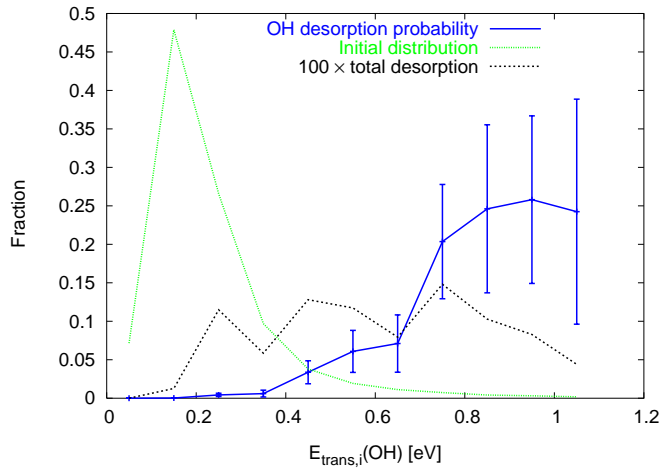}}
    \caption{The probabilities of OH desorption as functions of initial OH translational energy averaged over the
    three top monolayers of amorphous ice. Also
    shown are the initial distribution of OH translational energies and the total desorption probability
    (desorption probability $\times$ initial distrbution). The error bars correspond to a 95\% confidence interval.} 
    \label{FigOutvEtOH}
\end{figure}

Similarly, the desorption probability of OH has been plotted in Fig. \ref{FigOutvEtOH} as function of
initial OH translational energy in the top three monolayers. Also in this case the desorption probability increases with
increasing initial kinetic energy. The effect is much stronger than for the case of H atoms, which reflects the much
stronger binding energy of OH to its surroundings compared to that of the H atom. Interestingly, if the desorption 
probability is weighted with the initial distribution of translational energies, the total OH desorption probability (for the 
top three monolayers) increases rapidly at 0.2 eV (the average initial translational energy) and attains a basically
constant value of 0.1\% for all energies above that.

\begin{figure}[h]
    \caption{Probability of indirect H$_2$O desorption (``kick-out") as function of initial H atom
    translation energy (divided into bins of 0.5 eV intervals) averaged over the top six monolayers of amorphous ice.
    The error bars correspond to a 95\% confidence interval.}
\resizebox{\hsize}{!}{\includegraphics{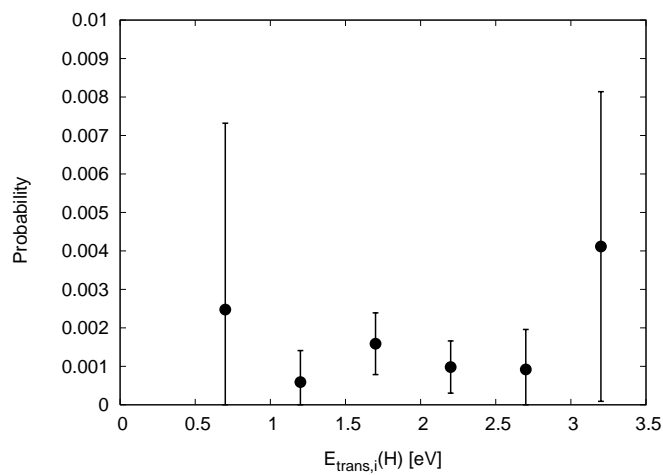}}
    \label{FigDesH2OvEtH}
\end{figure}

The dependence of the indirect H$_2$O desorption probability on the translational energy of the H atom
is found to be weak (see Fig. \ref{FigDesH2OvEtH}). There is not much evidence of any variation with
translational energy and the desorption probability is fairly constant at 0.1\% (averaged over the top six monolayers)
over the whole energy range. It would be natural to expect that there could be a strong dependence on translational
energy, but as discussed above the desorption is in most cases a combined effect of a repulsive force
from the excited molecule, a lowered binding energy, and the momentum transfer from the H atom. Apparently, this
allows also H atoms with relatively low translational energies to kick out H$_2$O molecules.

\subsubsection{Dependence on photon energy}
\label{DepPhoEn}

\begin{figure}[h]
   \caption{Probabilities of desorption of H atoms, OH radicals, and H$_2$O molecules
   as function of excitation energy averaged over the top six monolayers of amorphous ice. 
   The error bars correspond to a 95\% confidence interval.}
\resizebox{\hsize}{!}{\includegraphics{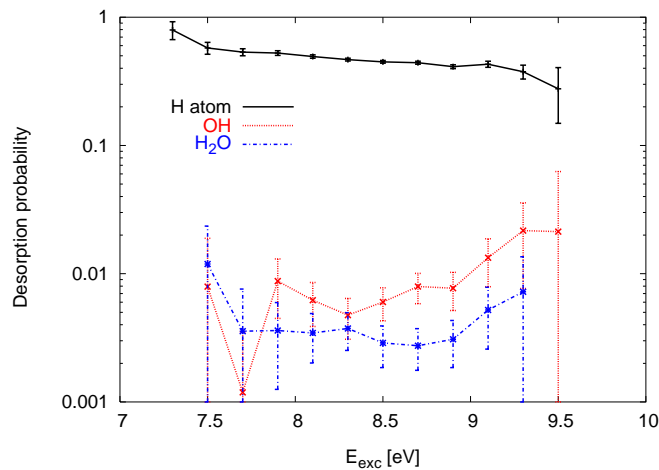}}
    \label{FigDesPvEexc}
\end{figure}

Considering the desorption probabilities as functions of the excitation energy (Fig. \ref{FigDesPvEexc})
it is interesting to note that
H atom desorption becomes {\em less} probable with increasing excitation energy (0.8 at 7.3 eV and 0.3 at 9.5 eV). 
Since it was shown that the average initial
translational energy mainly increases with increasing excitation energy (Fig. \ref{FigEtvvEexc})
and that the desorption probability increases with increasing
translational energy (Fig. \ref{FigOutvEtH}) this seems like a paradox.
The simple explanation of this behavior is that the lower excitation energies
dominate in the top monolayers while the more energetic UV photons are mainly absorbed towards the bulk of the ice
\citep[see, e.g., Fig. 5 of][]{and05}.
The desorption probability summed over the whole excitation energy range decreases rapidly with depth into the ice
(Table \ref{TabDes1}) and therefore this unexpected behavior is found. The desorption of OH seems to increase with
increasing excitation energy. This is a reflection of the fact that even though the {\em average} initial translational
energy of OH varies only slightly over the excitation energy range (Fig. \ref{FigEtvvEexc}), there is a high-energy tail
of the OH translational energy distribution (see Fig. \ref{FigOutvEtOH}) that becomes larger with higher excitation energies.
The desorption probability of H$_2$O does not show strong dependence on excitation energy, 
but considering the rather large error bars some energy dependence cannot be entirely ruled out.  

\subsection{Mobility of photoproducts}
\label{MobProd}

The H atoms produced
in the photodissociation event are found to be quite mobile in
the ice. On average the H atoms that become trapped move 8 {\AA}
from their original locations. In extreme cases distances over 70
{\AA} are recorded. The OH radicals formed in the ice move only about 1 {\AA}
with maximum distances moved of 5 {\AA}. However, OH radicals formed from
photodissociation in the top three monolayers are able in some cases
to move tens of {\AA} \emph{on top} of the surface (up to more than 60
{\AA}). The fact that some of the photofragments are able to move
large distances implies an increased probability of reactions with
other species in or on the ice, than if they would remain in the
immediate vicinity of their point of origin. For more details see
\citet{and06}.

\subsection{Comparison to experiments}
\label{CompExp}

\begin{figure}[h]
\resizebox{\hsize}{!}{\includegraphics{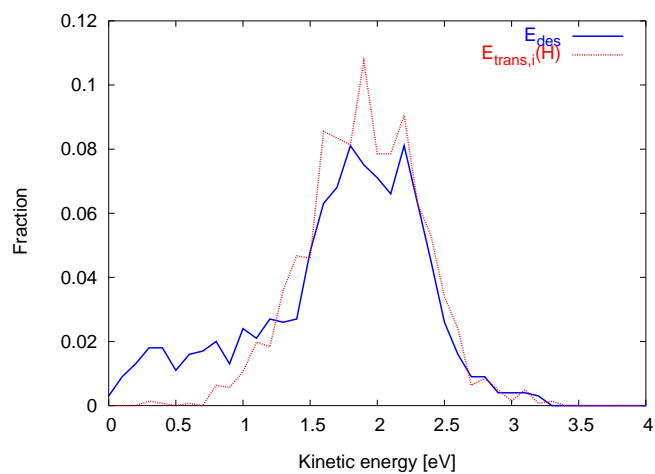}}
    \caption{Calculated distributions of H atom initial translational energies and desorption energies upon
    UV absorption at $E_{\mathrm{exc}}$ = 7.9 eV summed over the top six monolayers in amorphous ice.}
    \label{FigTrans79}
\end{figure}

In the experiments by \citet{yab06} on UV irradiation of polycrystalline and 
amorphous ices at 100 K absorption at $\lambda$ = 157 nm ($E_{\mathrm{exc}}$ = 7.9 eV)
was found to result in H atom desorption. 
The translational energy distribution of the desorbing atoms was observed to
consist of three components at 0.61 eV, 0.081 eV, and 0.014 eV, respectively.
The lowest-energy component seems to consist of atoms that have thermalized prior to desorption.
Thermal desorption occurs on a time scale that is likely much longer than would be feasible
to do with molecular dynamics simulations. Therefore, it is not likely that we would see 
this third component in our calculations, but the other two should be possible to reproduce. 
In Fig. \ref{FigTrans79} the calculated distributions of 
initial translational energies and desorption energies of
the released H atoms following excitation at 7.9 eV are shown. It is seen that the initial translational
energy has a peak at around 1.9 eV and the desorption energy peaks at about the same energy. This clearly is
much higher than found in the experiments, so it seems the energy of the desorbing H atoms is overestimated
in our calculations. There could be three explanations to this behavior: (i) either the H atoms lose more energy
prior to desorption or (ii) they are initially formed with less translational energy or (iii) a combination
of these two effects. The possible sources of loss of highly energetic H atoms in the ice that cannot be treated by
our calculations are: (a) the loss of energy by excitation of intramolecular modes in collisions with H$_2$O molecules
and (b) reactions with H$_2$O molecules to form, e.g., H$_2$ and OH. For a thorough discussion
see \citet{and06}. Since we cannot directly tell how effective mechanism (i) is it
is hard to speculate how much the desorption energy distribution is cooled through energy transfer and reaction.
It is easier to speculate about mechanism (ii), since it is possible to monitor the dependence of the average
desorption energy on the initial translational energy. If the initial translational energy
is around 1 eV the average desorption energies lie in the range 0.4--0.8 eV, which would be in much
better accord with the experimentally measured desorption energies. If this is the most important mechanism for 
cooling the desorption energy distriubution, then at $E_{\mathrm{exc}}$ = 7.9 eV the initial H atom translational energies
are overestimated by roughly 1 eV. As discussed in Sect. \ref{PhotoDesProb} this would probably have little importance
for the H$_2$O desorption probability (Fig. \ref{FigDesH2OvEtH}).
However, the desorption probability of H atoms could be somewhat lower
than predicted in our calculations, but it is still quite likely to be of the same
order of magnitude (Fig. \ref{FigOutvEtH}). If the average H atom translational energy is overestimated it
is also likely that the OH translational energy is somewhat overestimated. Considering the weak dependence on
excitation energy (Fig. \ref{FigEtvvEexc}) the {\em average} OH translational energy might not be highly overestimated,
but the high energy tail could be smaller, meaning that less OH radicals desorb than predicted here.
If the initial kinetic energies of the photofragments are overestimated, the only possibility to account for the
blue shift of the excitation energy is that the intermolecular repulsion is underestimated. This could have as an
interesting effect that the indirect desorption of the surrounding molecules could be {\em underestimated}, since
they would experience an even larger repulsive force from the excited molecule than predicted by our calculations.
This remains to be investigated.

The high probabilities of H atom desorption is also in accordance with the
finding of \citet{ger96} that their
UV-irradiated water ice was most likely depleted of H atoms, since a
large amount of oxygen rich products was found.
Our results on mobility of the released photofragments \citep[Sect. 3.3, ][]{and06} have been supported by recent
experiments by \citet{ell07}, who observed
an average separation of H and OH of 7$\pm$2 $\AA$ immediately following photodissociation of H$_2$O upon 
UV irradiation of liquid water. This is
in excellent agreement with our calculated value of 8 $\AA$ for the average distance from the site of
photodissociation of the H atoms in amorphous ice. Even though the temperatures are quite different
in the two cases, it is expected that liquid water and compact amorphous ice are quite similar on the
short time scales during which photodissociation takes place.

A direct comparison with the water ice photodesorption results by 
\citet{wes95a,wes95b} is difficult to make
since they used Lyman-$\alpha$ radiation, which leads to excitation to
a higher absorption band than considered here 
\citep{kob83}. Their findings of basically no photodesorption of intact H$_2$O at
low temperatures in the limit of single-photon absorption can
therefore be neither refuted nor confirmed by our results. However, the detection
of desorbing H$_2$ and O$_2$, and possibly OH and H$_2$O$_2$ actually agree with
the present results \citep{wes95a} (see below). 
In the experiments of \"Oberg et al. (submitted to ApJ) the photodesorbing material
is detected in the form of OH, H$_2$O, H$_2$, and O$_2$. This is the first time
a positive detection of OH photodesorption has been reported. The detection of OH
agrees nicely with our simulations (see also Sect. \ref{DiscAstro}).

It is important to bear in mind that even though
only H, OH, and H$_2$O would desorb upon absorption of a \emph{single} UV photon,
in the experimental setup a large amount of UV photons may be absorbed in the ice
during a relatively short time interval. This makes it possible to produce
significant amounts of H and OH in the ice, and possibly also O atoms from for
instance the photodissociation of the formed OH radicals. Given that thermal
diffusion is effective in moving these reactive species into close contact,
all the aforementioned desorbing species can be accounted for through recombination
followed by desorption.

\section{Discussion and astrophysical implications}
\label{DiscAstro}

Based on the results presented in this paper some important conclusions can be drawn on the possible
outcomes of UV irradiation of water ice in interstellar environments. First, it is important
to realize that by far the most likely species to desorb are H atoms, followed by OH radicals. In addition, when there
is removal of H$_2$O from the surface it seems likely that most of it comes off in the form of separate H and OH fragments.
Therefore, there is not a one-to-one correspondence between H$_2$O molecules removed from a surface and H$_2$O appearing in the gas phase.

So far, this paper has been concerned with probabilities of photoinduced processes following absorption of one UV photon in a specific
layer in the ice. However, not all incident UV photons are absorbed by molecules in the top six monolayers. To estimate 
desorption probabilities per \emph{incident} photon rather than per \emph{absorbed} photon one needs information
about the absorption cross section. In our semiclassical simulations we are not able to calculate the absolute absorption
cross section. However, \citet{mas06} have measured the absorption cross section of water ice at 25 K and found the peak absorption
cross section in the first absorption band to be about $6\times10^{-18}$ cm$^2$ at an excitation energy of 8.61 eV. This leads to an
absorption probability of 0.007 photons ML$^{-1}$ (see Appendix \ref{AppA} for an outline of this calculation).

From the above estimate of the absorption probability it is possible to calculate photodesorption probabilities per incident UV photon.
This is done by weighting the desorption probabilities per absorbed photon for each monolayer by the absorption probability
for the specific monolayer with the absorption probabilities in any upper monolayers subtracted from the incoming photon flux.
Using the information in Table \ref{TabDes2} one arrives at a probability of removal of H$_2$O from the ice of 
$3.7 \times 10^{-4}\ \mathrm{photon}^{-1}$. About 60\% of the removed H$_2$O comes off in the form of H + OH, 20\% desorbs as recombined
H$_2$O, and the remaining 20\% consists of H$_2$O ``kicked" out from the surface. The total photodesorption yield of intact H$_2$O
molecules is $1.4 \times 10^{-4}\ \mathrm{photon}^{-1}$. For OH desorption the probability is 
$3 \times 10^{-4}\ \mathrm{photon}^{-1}$,
which includes both desorption with and without H atoms. The H atom photodesorption probability 
is relatively high, $0.02\ \mathrm{photon}^{-1}$.
The above ratios of photodesorption of OH and H$_2$O are in excellent agreement with the recent experiments by \"Oberg et al. (submitted to ApJ),
which inferred that roughly equal amounts of OH and H$_2$O photodesorb at low surface temperature (18 K).
They obtain a total photodesorption yield of about $1.3 \times 10^{-3}$ in the low-temperature limit, which is about 3 times higher
than what is found from our simulations. Given the experimental uncertainties and the approximations made in the simulations this
can be considered as good agreement.
Commonly used estimates of H$_2$O
photodesorption probabilities in the range $1\times 10^{-4}$--$3.5\times 10^{-3}$ have been used to model different environments in
agreement with observations \citep{ber95,wil00,sne05,dom05,ber05}. Our results indicate that these estimates are reasonable.
However, the finding that less than half of the desorbed material leaves the grain in the form of intact H$_2$O molecules
is something that should be included in the models.

The results presented here are only for photoinduced processes that are followed until the photoproducts desorb or are thermalized within the
ice. This is all happening on a picosecond time scale. For longer time scales there is the possibility of thermal desorption of especially
the H atoms. These are relatively weakly bound to the ice surface and it is quite probable that some fraction of the released H atoms
desorb after thermalization in the ice. This contribution to the desorption probability is therefore not included in the above estimate.

It is interesting to discuss the possible effects of the overall morphology of the ice surface.
It is clear that the vast majority of water ice surfaces in the interstellar medium are amorphous \citep{hag81}. However, it is debated
whether the ice is mostly porous, as found in vapor deposited ice \citep{may86,kim01a,kim01b}, or more compact \citep{fra04,pal06}.
A complicating factor is that most likely the ice is formed through chemical reactions on grains rather than accretion of H$_2$O
molecules from the gas phase \citep{one99}. Therefore, the exact ice morphology that results from such a chemical build-up of the
ice is not yet clear (see however \citet{cup07}). If a porous ice surface is subjected to UV irradiation one could have release of H, OH, or H$_2$O into a void
in the ice rather than directly into the gas phase. This is inferred to happen in the experiments by \citet{yab06} where a large
component of the desorbing H atoms released after UV irradiaton of amorphous ice are thermalized, likely due to H atoms being accommodated 
within a void in the ice and then desorbing thermally. Some OH and H$_2$O could photodesorb in a similar way. However, if the
path to reach the gas phase from inside a pore is restricted, these species have a high probability of being trapped inside
a pore because of their strong attractive interactions with the ice. In practice, the OH and H$_2$O released in this
way might not show up in desorption. Therefore, if a porous ice is considered it is necessary to distinguish
between the surface area that is directly exposed to the gas phase and that which is within a void with restricted access to the outside.
In conclusion, the H atom desorption from a porous ice is likely to be different from that of compact non-porous ice, but
the desorption yields of OH and H$_2$O are not necessarily very different in the two types of ice. 

The cosmic ray induced UV flux inside dense clouds is about $10^4\ \mathrm{cm}^{-2}\mathrm{s}^{-1}$ \citep{she04}. 
For an ice-coated grain with a typical 
size of 0.1 $\mu$m, this would give an arrival rate of about 1 UV photon per day. The case of a single UV absorption
event as described in our simulations therefore gives a realistic picture of photodesorption in dense clouds. In the laboratory, the UV
flux is many orders of magnitude higher and multiple absorptions of UV photons within the ice surface in a short time interval may
drive secondary reactions of photodissociation products from different H$_2$O molecules. Experiments with different UV flux levels
down to low levels will be needed to provide quantitative data on photodesorption yields relevant for astrophysical applications. 

As has also been noted in our previous work another important aspect is the release of reactive species into the ice. That would
have implications for reactivity in the ice with, e.g., CO that could react with energetic H or OH to form HCO or CO$_2$. This could be
one clue to unraveling the mystery of CH$_3$OH and CO$_2$ formation in the interstellar medium. Indeed, formation of CO$_2$ is readily observed
when a mixed H$_2$O:CO ice is photolysed \citep{dhe86,wat02,wat07}.

\section{Conclusions}

We have shown that it is possible to have H$_2$O photodesorption upon UV absorption to the first absorption band 
in the top five monolayers of an amorphous ice surface.
The main mechanisms for this photodesorption are either photodissociation followed by recombination of H and OH and subsequent desorption
of the recombined H$_2$O molecule or a ``kick-out" of another H$_2$O molecule in the ice by the energetic H atom released from photodissociation
or, less likely, by the energy released from a recombined H$_2$O molecule. In most cases, however, removal of an H$_2$O molecule from the ice is
in the form of separate H and OH fragments. An estimate of the photodesorption yield per incident UV photon 
from our calculations agrees well with the H$_2$O photodesorption yields that are commonly used in modeling astrophysical environments.

UV absorption leads in most cases to desorption of H atoms or the trapping of H and OH in the ice either as separate fragments or as recombined H$_2$O.
The desorption of H atoms is about 2 or 3 orders of magnitude more probable than desorption of OH and H$_2$O. The OH desorption probability is about twice the H$_2$O
desorption probability. The high mobility of H atoms inside the ice and OH radicals on the ice surface will facilitate formation of other
molecules such as CO$_2$.

\begin{acknowledgements}
We thank Karin \"Oberg, Herma Cuppen, and Geert-Jan Kroes for stimulating discussions.
Some of the calculations reported here were performed at Chalmers Centre for Computational
Science and Engineering (C3SE) computing resources. This research was funded by a Netherlands Organization for Scientific Research (NWO) Spinoza
grant [for one of the authors (E.F.v.D)] and a NWO-CW Top grant.

\end{acknowledgements}

\begin{appendix}
\section{Estimate of absorption probability}
\label{AppA}
To calculate an absorption probability per monolayer in an interstellar ice surface one needs
an estimate of the effective area taken up by one molecule. Since the angle of incidence of the photon is arbitrary,
all possible incidence angles have to be taken into account, not only normal incidence.
The surface area of our simulation cell is 
$22.4\ \mathrm{\AA}\ \times\ 23.5\ \mathrm{\AA}\ =\ 526\ \mathrm{\AA}^2\ =\ 5.26\ \times10^{-14}\ \mathrm{cm}^2$.
If one assumes a flat surface, the average effective surface area seen by a photon from any incidence angle between 0$^{\circ}$ (normal incidence)
and 90$^{\circ}$ (parallel to the surface) is the actual surface area divided by two. This is arrived upon by
the following expression:

\begin{equation}
\langle A_\mathrm{eff} \rangle = \frac{\int^{\pi / 2}_0 A \cos\theta \sin\theta d\theta}
{\int^{\pi / 2}_0 \sin\theta d\theta} = \frac{A}{2},
\end{equation}

with the effective area given by $A_\mathrm{eff} = A\cos\theta$ where $A$ is the actual surface area and $\theta$
is the angle of incidence. The average effective surface area of the simulation cell is 2.63$\times$10$^{-14}$ cm$^2$.
Each monolayer consists of 30 H$_2$O molecules. A single molecule will therefore have an average effective area 
$\langle A_\mathrm{eff}^\mathrm{mol} \rangle = 8.77\times10^{-16}\ \mathrm{cm}^2$.
\citet{mas06} measured the peak absorption cross section, $\sigma$, in the first absorption band to be about 6$\times$10$^{-18}$ cm$^2$
around an excitation energy of 8.61 eV.
With this value the absorption probability per monolayer becomes 
$P_\mathrm{abs}^\mathrm{ML} = \sigma / \langle A_\mathrm{eff}^\mathrm{mol} \rangle = 7\times10^{-3}$ (this number may vary somewhat with
excitation energy). If one
considers an infinitely deep ice surface all photons will be absorbed within the ice.
For an ice surface in the interstellar medium consisting of a finite number of layers, 
the value of $P_\mathrm{abs}^\mathrm{ML}$ may vary depending on the shape of the grain,
since the surface is not necessarily flat. However, it is likely that any photon that passes through an ice mantle will be
absorbed by the silicate grain, leading to almost total absorption of the incident photons within the grain.

\end{appendix}

\end{document}